\documentclass[12pt]{article}
\usepackage[dvips]{color}
\usepackage{epsfig}
\usepackage{amsmath}
\usepackage{graphicx}
\def\Box{\hbox{$\rlap{$\sqcup$}\sqcap$}}
\begin{document}
\title{\bf Lorentz Force and Ponderomotive Force in the Presence of a Minimal Length }

\author{B. Khosropour \thanks{E-mail: $b _ -khosropour@kazerunsfu.ac.ir$}\hspace{1mm}\\
 {\small {\em  Department of Physics, Faculty of Sciences,
Salman Farsi University of Kazerun , Kazerun ,73175-457, Iran}}\\
\\}
\date{\small{}}
\maketitle
\begin{abstract}
In this work, according to the electromagnetic field tensor in the
framework of generalized uncertainty principle (GUP), we obtain the
Lorentz force and Faraday's law of induction in the presence of a
minimal length. Also, the ponderomotive force and ponderomotive
pressure in the presence of a measurable minimal length are found.
It is shown that in the limit $\beta\rightarrow0$, the generalized
Lorentz force and ponderomotive force become the usual forms. The
upper bound on the isotropic minimal length is estimated.

\noindent
\hspace{0.35cm}



\end{abstract}
\newpage

\section{Introduction}
Different theories of quantum gravity such as the loop quantum
gravity, string theory and noncommutative geometry have been
proposed for exploring the unification between the general theory of
relativity and the standard model of particle physics [1]. Although
these theories are different in concepts, all of these ivestigations
lead to unique belief which predicts the existence of a minimal
length scale. An immediate consequence of existence of a minimal
length gave rise to the modification of Heisenberg uncertainty
principle. Nowadays the modified uncertainty principle is known
generalized uncertainty principle (GUP) [2]. The generalized
uncertainty principle (GUP) can be written as follows:
\begin{equation}
\triangle X \triangle P\geq\frac{\hbar}{2}\left[1+\beta(\Delta
P)^{2}\right],
\end{equation}
where $\beta$ is a positive parameter [3,4] and it is obvious that
$\triangle X $ is always greater than $(\triangle
X)_{min}=\hbar\sqrt{\beta}$. In the recent years, many studies have
been devoted to the gravity and reformulation of quantum field
theory and electrodynamics in the presence of a minimal length scale
[5-20]. Quesne and Tkachuk have introduced a Lorentz covariant
deformed algebra which describes a $D+1$-dimensional space time is
characterized by the following deformed commutation relations:
\begin{eqnarray}
\left[X^{\mu},P^{\nu}\right] &=&-i\hbar[(1-\beta
P_{\rho}P^{\rho})g^{\mu\nu}-\beta'P^{\mu}P^{\nu}], \\
\nonumber \left[X^{\mu},X^{\nu}\right] &=& i\hbar\
\frac{2\beta-\beta'-(2\beta+\beta')\beta P_{\rho}P^{\rho} }{1-\beta
P_{\rho}P^{\rho}}(P^{\mu}X^{\nu}-P^{\nu}X^{\mu}), \\
\nonumber
\left[P^{\mu},P^{\nu}\right] &=& 0,
\end{eqnarray}
where $\mu,\; \nu, \; \rho=0,1,2,\cdots,D$ and $\beta ,\; \beta'$
are two positive deformation parameters. In Eq. (2), $X^{\mu}$ and
$P^{\mu}$ are position and momentum operators in the framework of
GUP and  $g_{\mu\nu}=g^{\mu\nu}=diag(1,-1,-1,\cdots,-1)$. An
immediate consequence of Eq. (2) is the following an isotropic
minimal length
\begin{equation}
(\triangle X^{i})_{min}=\hbar\sqrt{(D\beta+\beta')
[1-\beta\langle(P^{0})^{2}\rangle]}\quad , \quad\forall i\in \{1,2,
\cdots ,D\}.
\end{equation}
In the present work, we study the Lorentz force and ponderomotive
force in the presence of a minimal length. The paper is organized as
follows: In Sec. 2, the lorentz force and Faraday's law of induction
is obtained in the presence of a minimal length. In Sec. 3, the
ponderomotive force in the framework of GUP is found and the
relative modification of ponderomotive pressure is obtained. Also,
the upper bound on the isotropic minimal length is estimated. Our
conclusions are presented in Sec. 4.
\section{Lorentz Force in the Presence of a Minimal Length}
The aim of this section is finding the Lorentz force and Faraday's
law in the presence of a minimal length based on the Quesene-Tkachuk
algebra. Hence, we need a representation which satisfies the
generalized commutation relations in Eq. (2). In Ref. [23], Tkachuk
introduced the following representation which satisfies the deformed
algebra up to the first order in deformation parameters $\beta$ and
$\beta'$
\begin{eqnarray}
X^{\mu} &=& x^{\mu}- \frac{2\beta-\beta'}{4}(x^{\mu}
p_{\rho}p^{\rho}+p_{\rho}p^{\rho}x^{\mu}), \\ \nonumber
P^{\mu} &=&
(1-\frac{\beta'}{2}p_{\rho}p^{\rho})p^{\mu},
\end{eqnarray}
where $x^{\mu}$ and $p^{\mu}=i\hbar \frac{\partial}{\partial
x_{\mu}}=i\hbar\partial^{\mu} $ are position and momentum operators
in usual quantum mechanics. It is interesting to note that in the
special case of $\beta'=2\beta$, the position operators commute to
the first order in deformation parameter $\beta$, i.e.,
$[X^{\mu},X^{\nu}]=0$. In this special case, the Quesene-Tkachuk
algebra becomes
\begin{eqnarray}
\left[X^{\mu},P^{\nu}\right] &=&-i\hbar[(1-\beta P_{\rho}P^{\rho})g^{\mu\nu}-2\beta P^{\mu}P^{\nu}],
\\ \nonumber
\left[X^{\mu},X^{\nu}\right] &=& 0, \\
\nonumber
\left[P^{\mu},P^{\nu}\right] &=& 0.
\end{eqnarray}
The following representations satisfy Eq. (5), in the first order in
$\beta$:
\begin{eqnarray}
X^{\mu} &=& x^{\mu}, \\
P^{\mu} &=& (1-\beta p_{\rho}p^{\rho})p^{\mu}.
\end{eqnarray}

\subsection{The Modified Lorentz Force }
The Lorentz force is the combination of electric and magnetic force
on a charged particle due to electromagnetic fields. If we consider
a particle of charge $q$ moves with velocity $\textbf{v}$ in the
presence of an electric field $\textbf{E}$ and a magnetic field
$\textbf{B}$, then the Lorentz force is given by [24]
\begin{eqnarray}
\frac{d\textbf{p}}{dt}=q(\textbf{E}+\textbf{V}\times\textbf{B}).
\end{eqnarray}
We know that $\textbf{p}$ transforms as the space part of the
momentum four-vector,
\begin{eqnarray}
p^{\mu}=(p^{0},\textbf{p})=(\frac{E}{c},\textbf{p})=m(U^{0},
\textbf{U}),
\end{eqnarray}
and $U^{\alpha}$ is the velocity four-vector. The covariant form of
the Lorentz force in a $3+1$-dimensional space time can be written
as follows[24]
\begin{eqnarray}
\frac{dp^{\alpha}}{d\tau}=qU_{\rho}F^{\alpha \rho},
\end{eqnarray}
where $F^{\alpha \rho}$ is the electromagnetic field tensor and
$\tau$ is the proper time which are the following definition
\begin{eqnarray}
F^{\alpha
\rho}=\partial^{\alpha}A^{\rho}-\partial^{\rho}A^{\alpha},\\
\tau=\frac{t}{\gamma}=\frac{t}{\sqrt{1-\frac{v^{2}}{c^{2}}}}.
\end{eqnarray}
In a $3+1$ dimensional space time, the components of the
electromagnetic field tensor $F^{\alpha \rho}$ can be written as
\begin{eqnarray}
F^{\alpha\rho}&=& \left( {\begin{array}{cccc}
   0 & -{E_{x}}/{c}\ & -{E_{y}}/{c}  & -{E_{z}}/{c}  \\
   {E_{x}}/{c} & 0 & -B_{z} &B _{y} \\
   {E_{y}}/{c} & B_{z} & 0 & -B_{x} \\
   {E_{z}}/{c} & -B_{y} & B_{x} & 0
    \end{array}} \right).
\end{eqnarray}
Now, we obtain the electromagnetic field tensor in the presence of a
minimal length based on the Quesene-Tkachuk algebra. For this
purpose, let us write the electromagnetic field tensor in Eq. (11)
by using the representations Eqs. (6) and (7), that is
\begin{eqnarray}
x^{\mu}\longrightarrow  X^{\mu}&=&x^{\mu}, \\ \nonumber
\partial^{\mu}\longrightarrow D^{\mu}&:=&(1+\beta\hbar^{2}\Box)\partial^{\mu},
\end{eqnarray}
where $\Box:=\partial_{\mu}\partial^{\mu}$ is the d'Alembertian
operator. If we substitute Eq. (14) into Eq. (11), we will obtain
the following modified electromagnetic field tensor
\begin{eqnarray}
{\cal
F}^{\alpha\rho}&=&D^{\alpha}A^{\rho}-D^{\rho}A^{\alpha}=(1+\beta\hbar^{2}\Box)\partial^{\alpha}A^{\rho}-(1+\beta\hbar^{2}\Box)\partial^{\rho}A^{\alpha}\\
\nonumber &=&F^{\alpha\rho}+\beta\hbar^{2}\Box
F^{\alpha\rho}=(1+\beta\hbar^{2}\Box)F^{\alpha\rho}.
\end{eqnarray}
The term $\beta\hbar^{2}\Box F^{\alpha\rho}$ in Eq. (15) can be
considered as a minimal length effect. By inserting the modified
electromagnetic field tensor into Eq. (10), the modified Lorentz
force can be obtained as follows
\begin{eqnarray}
\frac{dp^{\alpha}}{d\tau}=q U_{\rho} {\cal
F}^{\alpha\rho}=qU_{\rho}(1+\beta\hbar^{2}\Box)F^{\alpha\rho}.
\end{eqnarray}
According to Eq. (13), Eq. (15) can be written in the vector form as
follows
\begin{eqnarray}
\frac{d\textbf{p}}{dt}&=&q[\textbf{E}(\textbf{x},t)+\textbf{v}\times\textbf{B}(\textbf{x},t)]\\
\nonumber &+&q\beta
\hbar^{2}(\frac{1}{c^{2}}\frac{\partial^{2}}{\partial
t^{2}}-\nabla^{2})[\textbf{E}(\textbf{x},t)+\textbf{v}\times\textbf{B}(\textbf{x},t)].
\end{eqnarray}
In the limit $\beta\rightarrow 0$, the generalized Lorentz force in
Eq. (16) becomes the usual Lorentz force.

\subsection{Faraday's Law of Induction in the Presence of a Minimal Length}
If we consider a circuit wire in the presence of a magnetic field,
we find the induced electromotive force in the wire. The induced
electromotive force around the circuit wire is equal to Faraday's
law of induction, that is
\begin{eqnarray}
\epsilon=-\frac{d\phi_{B}}{dt},
\end{eqnarray}
where $\phi_{B}$ is the magnetic flux through the wire. It is easy
to derive the Faraday's law from the Lorentz force and the Maxwell
equations. As we know that the electric field and the induced
electromotive force are defined by
\begin{eqnarray}
\textbf{E}(\textbf{x},t)=\frac{\textbf{F}(\textbf{x},t)}{q},\\
\epsilon=\int_{c}\textbf{E}.d\textbf{L}.
\end{eqnarray}
If we substitute Eq. (19) into Eq. (20) and considering Lorentz
force in Eq. (8), we have
\begin{eqnarray}
\epsilon&=&\int_{c}\textbf{E}.d\textbf{L}=\int_{c}\frac{\textbf{F}(\textbf{x},t)}{q}\\
 \nonumber
 &=&\int_{c}\textbf{E}.d\textbf{L}+\int_{c}[\textbf{v}\times\textbf{B}(\textbf{x},t)].d\textbf{L}.
\end{eqnarray}
Hence, if we use the Kelvin-Stokes theorem and considering the
Maxwell-Faraday's equation ($\nabla\times\textbf{E}=-\frac{\partial
\textbf{B}}{\partial t},$) , we will obtain
\begin{eqnarray}
\epsilon=-\int\int_{S}d\textbf{s}\cdot\frac{\partial
\textbf{B}(\textbf{x},t)}{\partial
t}+\int_{c}(\textbf{v}\times\textbf{B}(\textbf{x},t))\cdot
d\textbf{L}.
\end{eqnarray}
Now, using the Leibniz integral rule and the $\nabla\cdot
\textbf{B}=0$, we can find the following Faraday's law of induction
\begin{eqnarray}
\epsilon=-\frac{d}{dt}\int\int_{S}d\textbf{s}\cdot\textbf{B}(\textbf{x},t)=-\frac{d\phi_{B}}{dt}.
\end{eqnarray}
Let us find the Faraday's law in the presence of a minimal length.
If we use Eqs. (17) and (21), we can write the modified induced
electromotive force as follows

\begin{eqnarray}
\epsilon_{ML}&=&\int_{c}\frac{\textbf{F}_{ML}}{q}
=[\int_{c}\textbf{E}.d\textbf{L}+\int_{c}(\textbf{v}\times\textbf{B})\cdot
d\textbf{L}]\\ \nonumber&+&\beta
\hbar^{2}(\frac{1}{c^{2}}\frac{\partial^{2}}{\partial
t^{2}}-\nabla^{2})[\int_{c}\textbf{E}.d\textbf{L}+\int_{c}(\textbf{v}\times\textbf{B})\cdot
d\textbf{L}].
\end{eqnarray}
According to our previous work[11], we have found that the
homogeneous Maxwell's equations
$(\nabla\times\textbf{E}=-\frac{\partial \textbf{B}}{\partial t},
\nabla\cdot\textbf{B}=0)$ were not modified by the effect of minimal
length. Therefore by using the Kelvin-Stokes theorem and
Maxwell-Faraday, Eq. (24) will be became
\begin{eqnarray}
\epsilon_{ML}&=& =[-\int\int_{S}d\textbf{s}.\frac{\partial
\textbf{B}(\textbf{x},t)}{\partial t}+\int_{c}(\textbf{v}\times\textbf{B}(\textbf{x},t))\cdot d\textbf{L}]\\
\nonumber&+&\beta
\hbar^{2}(\frac{1}{c^{2}}\frac{\partial^{2}}{\partial
t^{2}}-\nabla^{2})[-\int\int_{S}d\textbf{s}.\frac{\partial
\textbf{B}(\textbf{x},t)}{\partial
t}+\int_{c}(\textbf{v}\times\textbf{B}(\textbf{x},t))\cdot
d\textbf{L}].
\end{eqnarray}
Now, by using the Leibniz integral rule and
$\nabla\cdot\textbf{B}=0$, the Faraday's law of induction in the
presence of a minimal length can be obtained as follows
\begin{eqnarray}
\epsilon_{ML}&=&[-\frac{d}{dt}\int\int_{S}d\textbf{s}\cdot\textbf{B}(\textbf{x},t)]+(\hbar\sqrt{\beta})^{2}(\frac{1}{c^{2}}\frac{\partial^{2}}{\partial
t^{2}}-\nabla^{2})[-\frac{d}{dt}\int\int_{S}d\textbf{s}\cdot\textbf{B}(\textbf{x},t)]\\
\nonumber
&=&[-\frac{d\phi_{B}}{dt}]+(\hbar\sqrt{\beta})^{2}(\frac{1}{c^{2}}\frac{\partial^{2}}{\partial
t^{2}}-\nabla^{2})[-\frac{d\phi_{B}}{dt}].
\end{eqnarray}

\section{Ponderomotive Force in the Presence of a Minimal Length}
Light waves exert a radiation pressure which is usually very weak
and hard to detect. When high-powered microwaves or laser beams are
used to confine plasmas, however, the radiation pressure reaches to
very high value [25]. When high-powered microwaves applied to a
plasma, this force is coupled to the particles in a somewhat subtle
way and is known the ponderomotive force. Many nonlinear phenomena
have a simple explanation in terms of the ponderomotive force. For
deriving this nonlinear force is to consider the motion of an
electron in the oscillating $\textbf{E}$ and $\textbf{B}$ fields of
a wave. By vanishing dc $\textbf{E}_{0}$ and $\textbf{B}_{0}$
fields, the electron equation of motion is [25]
\begin{eqnarray}
m\frac{d\textbf{v}}{dt}=-e(\textbf{E}(\textbf{r})+\textbf{v}\times\textbf{B}(\textbf{r})).
\end{eqnarray}
The nonlinearity comes partly from the $\textbf{v}\times\textbf{B}$
term, hence the term is no larger than
$\textbf{v}_{1}\times\textbf{B}_{1}$, where $\textbf{v}_{1}$ and
$\textbf{B}_{1}$ are the linear-theory values. Another part of the
nonlinearity, comes from evaluating $\textbf{E}$ at the actual
position of the particle rather than its initial position. By
considering a wave electric field of the form
\begin{eqnarray}
E=E_{s}(r)\cos(\omega t),
\end{eqnarray}
where $E_{s}(r)$ contains the spatial dependence and also in first
order, we may vanish the $\textbf{v}\times\textbf{B}$ term in Eq.
(27), therefore we have
\begin{eqnarray*}
m \frac{d\textbf{v}_{1}}{dt}=-e\textbf{E}(r_{0})
\end{eqnarray*}
and then
\begin{eqnarray}
\textbf{v}_{1}&=&\frac{-e}{m\omega}E_{s}\sin(\omega t),\\
\delta(\textbf{r}_{1})&=&\frac{e}{m\omega^{2}}E_{s}\cos(\omega t).
\end{eqnarray}
For finding the second order, we expand $\textbf{E}(\textbf{r})$
about $r_{0}$
\begin{eqnarray}
\textbf{E}(\textbf{r})=\textbf{E}(\textbf{r}_{0})+(\delta\textbf{r}_{1}\cdot\nabla)\textbf{E}|_{r=r_{0}}+\ldots,
\end{eqnarray}
and from Maxwell's equation, we can obtain $\textbf{B}_{1}$ as
follows
\begin{eqnarray}
\textbf{B}_{1}=-\frac{1}{\omega}\nabla\times\textbf{E}_{s}|_{r=r_{0}}\sin(\omega
t).
\end{eqnarray}
Now, we are ready to study the second order part of Eq. (27), hence
we have
\begin{eqnarray}
m
\frac{d\textbf{v}_{2}}{dt}=-e[(\delta\textbf{r}_{1}\cdot\nabla)\textbf{E}+\textbf{v}_{1}\times\textbf{B}_{1}].
\end{eqnarray}
Inserting Eqs. (29), (30) and (32) into Eq. (33) and averaging time,
we can find the ponderomotive force on a single electron as follows
\begin{eqnarray}
f=-\frac{1}{4}\frac{e^{2}}{m\omega^{2}}\nabla|E_{s}^{2}|.
\end{eqnarray}
Here we used $\langle \sin^{2}(\omega t)\rangle=\langle
\cos^{2}(\omega t)\rangle=\frac{1}{2}$.\\
Now, let us obtain the ponderomotive force in the presence of a
minimal length. According to Eq. (17), the electron equation of
motion in the presence of a minimal length is given by
\begin{eqnarray}
m\frac{d\textbf{v}}{dt}&=&-e[\textbf{E}(\textbf{r})+\textbf{v}\times\textbf{B}(\textbf{r})]\\
\nonumber
&-&e\beta\hbar^{2}(\frac{1}{c^{2}}\frac{\partial^{2}}{\partial
t^{2}}-\nabla^{2})[\textbf{E}(\textbf{r})+\textbf{v}\times\textbf{B}(\textbf{r})].
\end{eqnarray}
From Eq. (33), the modified second order part of Eq. (35) can be
written as follows
\begin{eqnarray}
m
\frac{d\textbf{v}_{2}}{dt}&=&-e[(\delta\textbf{r}_{1}\cdot\nabla)\textbf{E}+\textbf{v}_{1}\times\textbf{B}_{1}]\\
\nonumber
&-&e\beta\hbar^{2}(\frac{1}{c^{2}}\frac{\partial^{2}}{\partial
t^{2}}-\nabla^{2})[(\delta\textbf{r}_{1}\cdot\nabla)\textbf{E}+\textbf{v}_{1}\times\textbf{B}_{1}].
\end{eqnarray}
By inserting Eqs. (29), (30) and (32) into Eq. (36), we will obtain
\begin{eqnarray}
m
\frac{d\textbf{v}_{2}}{dt}&=&-\frac{e^{2}}{m\omega^{2}}[(\textbf{E}_{s}\cdot\nabla)\textbf{E}_{s}(\cos^{2}(\omega t))+(\textbf{E}_{s}\times\nabla\times\textbf{E}_{s})\sin^{2}(\omega t)]\\
\nonumber
&-&\frac{e^{2}}{m\omega^{2}}\beta\hbar^{2}(\frac{1}{c^{2}}\frac{\partial^{2}}{\partial
t^{2}}-\nabla^{2})[(\textbf{E}_{s}\cdot\nabla)\textbf{E}_{s}(\cos^{2}(\omega
t))+(\textbf{E}_{s}\times\nabla\times\textbf{E}_{s})\sin^{2}(\omega
t)].
\end{eqnarray}
After simplifying and averaging time, we can obtain the following
generalized ponderomotive force on a single electron
\begin{eqnarray}
\langle
f\rangle_{ML}=-\frac{1}{4}\frac{e^{2}}{m\omega^{2}}[\nabla|E_{s}^{2}|-((\hbar\sqrt{\beta})^{2}\nabla^{2})\nabla|E_{s}^{2}|].
\end{eqnarray}
It should be noted that in the limit $\beta\rightarrow0$, the
generalized ponderomotive force in Eq. (38) becomes the usual
ponderomotive force. Another hand, if we substitute $\beta'=2\beta$
into Eq. (3), we will find the following isotropic minimal length up
to the first order over the deformation parameter $\beta$
\begin{eqnarray}
(\triangle X^{i})_{min}=\hbar\sqrt{(D+2)\beta }\quad , \quad\forall
i\in \{1,2, \cdots ,D\}.
\end{eqnarray}
The isotropic minimal length in three spatial dimensions is given by
\begin{eqnarray}
(\triangle X)_{min}=\hbar\sqrt{5\beta }.
\end{eqnarray}
Here, by substituting Eq. (40) into Eq. (38), we have
\begin{eqnarray}
f_{ML}=-\frac{1}{4}\frac{e^{2}}{m\omega^{2}}[\nabla|E_{s}^{2}|-(\frac{(\triangle
X)_{min}^{2}}{5}\nabla^{2})\nabla|E_{s}^{2}|].
\end{eqnarray}
The force per $m^{3}$ is $f$ times the electron density $n_{0}$,
which can be defined in terms of plasma frequency $(\omega_{p})$ as
follows
\begin{eqnarray}
\omega_{p}=\sqrt{\frac{4\pi n_{0}e^{2}}{m_{e}}}.
\end{eqnarray}
From Eqs. (41) and (42) and since $E_{s}^{2}=2\langle E^{2}\rangle$,
we finally have for the generalized ponderomotive force the formula
\begin{eqnarray}
F_{ML}=-\frac{\omega_{p}^{2}}{\omega^{2}8\pi}[\nabla\langle
E^{2}\rangle-(\frac{(\triangle X)_{min}^{2}}{5}\nabla^{2})\nabla
\langle E^{2}\rangle].
\end{eqnarray}
Also, as we know the ponderomotive pressure has the following
definition
\begin{eqnarray}
F=-\nabla {\cal P}_{pond}.
\end{eqnarray}
According to Eq. (44), we can obtain the generalized ponderomotive
pressure as follows
\begin{eqnarray}
{\cal P}_{pond}=\frac{\omega_{p}^{2}}{\omega^{2}8\pi}[\langle
E^{2}\rangle-(\frac{(\triangle X)_{min}^{2}}{5}\nabla^{2}) \langle
E^{2}\rangle].
\end{eqnarray}

If we consider the first term $({\cal P}_{pond})_{0}$, the usual
ponderomotive pressure and the second term is its correction due to
the considered minimal length effect$({\cal P}_{pond})_{ML}$, the
relative modification of ponderomotive pressure can be found as
\begin{eqnarray}
\triangle{\cal
P}_{pond}=\frac{\omega_{p}^{2}}{\omega^{2}40\pi}[((\triangle
X)_{min}^{2}\nabla^{2}) (\langle E^{2}\rangle)].
\end{eqnarray}
Now we can estimate the upper bound on the isotropic minimal length
in the modified ponderomotive pressure. If we consider the value of
ponderomotive pressure is about $10^{6}Pa$ and also assuming
$\omega_{p}\approx10^{28}Hz$, $\omega\approx10^{13}Hz$. Therefore
from Eq. (46), we can estimate the following upper bound on the the
isotropic minimal length
\begin{eqnarray}
10^{6}&\approx& 10^{30}[((\triangle X)_{min}^{2}\nabla^{2}) (\langle
E^{2}\rangle)],\\ \nonumber (\triangle X)_{min}&\leq&10^{-12}m.
\end{eqnarray}
It is interesting to note that the estimation on the isotropic
minimal length is near to the minimal observable distance which was
proposed by Heisenberg($L\sim10^{-15}m$)[26,27]. On the other hand,
Nouicer has investigated the Casimir effect in the frame work of GUP
and obtains $(\triangle X^{i})_{min}\leq 15\times10^{-9}m$. The
upper bound on the isotropic minimal length in Eq. (47) is
compatible with the results of Refs. [30] and [31].\\
Although the above value for the minimal length $((\triangle
X^{i})_{min}\simeq 10^{-12}m)$ is about six orders of magnitude
larger than the electroweak length scale $(10^{-18}m)$, this value
is near to the reduced Compton wave length of electron
$(\lambda_{c}=3.86\times10^{-13}m)$.

\section{Conclusions}
In the last decade, many investigations have been done to compute
the corrections of different phenomena of electrodynamics and
quantum mechanics in the framework of GUP. According to these
studies, we can find that the implications of GUP can be sought at
atomic scales [28,29]. In this study, first we have obtained the
Lorentz force in the presence of a minimal length based on the
Quesne-Tkachuk algebra. Then from the generalized Lorentz force we
have found the Faraday's law of induction in the framework of GUP.
As many nonlinear phenomena have a explanation in terms of the
ponderomotive force, we have investigated the ponderomotive force in
the presence of a minimal length scale. Also, the generalized
ponderomotive pressure has been obtained. It should be emphasized
that in the limit $\beta\rightarrow0$, all of the generalized
lorentz force and Faraday's law and ponderomotive force become the
ordinary forms. Finally, the upper bound on the isotropic minimal
length scale has been estimated by using the experimental value of
ponderomotive pressure.It is interesting to note that the estimation
on the isotropic minimal length was near to the minimal observable
distance which was proposed by Heisenberg($L\sim10^{-15}m$).

\section*{Acknowledgments}
 We would like to thank the referees for their careful reading and constructive comments.

\end{document}